\documentclass[runningheads]{llncs} 
\usepackage[T1]{fontenc}
\usepackage{graphicx}
\usepackage{cite}
\usepackage{algorithm2e}
\usepackage{multirow}
\usepackage[table,xcdraw]{xcolor}
\usepackage{adjustbox}
\usepackage{cprotect}
\usepackage{color}
\usepackage[hyphens]{url} 
\usepackage{hyperref}

\begin{document}
\title{Layout-Aware Text Editing for Efficient Transformation of Academic PDFs to Markdown}
\titlerunning{Layout-Aware Text Editing for PDF-to-Markdown Transformation}
\author{Changxu Duan\orcidID{0000-0003-0547-0901}} 
\institute{Technische Universität Darmstadt, Darmstadt, Germany \\
\email{duan@linglit.tu-darmstadt.de}}
\authorrunning{C. Duan}
\maketitle              
\begin{abstract}
Academic documents stored in PDF format can be transformed into plain text structured markup languages to enhance accessibility and enable scalable digital library workflows.
Markup languages allow for easier updates and customization, making academic content more adaptable and accessible to diverse usage, such as linguistic corpus compilation.
Such documents, typically delivered in PDF format, contain complex elements including mathematical formulas, figures, headers, and tables, as well as densely layouted text.
Existing end-to-end decoder transformer models can transform screenshots of documents into markup language.
However, these models exhibit significant inefficiencies; their token-by-token decoding from scratch wastes a lot of inference steps in regenerating dense text that could be directly copied from PDF files.
To solve this problem, we introduce EditTrans, a hybrid editing-generation model whose features allow identifying a queue of to-be-edited text from a PDF before starting to generate markup language.
EditTrans contains a lightweight classifier fine-tuned from a Document Layout Analysis model on 162,127 pages of documents from arXiv.
In our evaluations, EditTrans reduced the transformation latency up to 44.5\% compared to end-to-end decoder transformer models, while maintaining transformation quality.
Our code and reproducible dataset production scripts are open-sourced.

\end{abstract}

\section{Introduction}
Transforming academic documents from PDF to markup languages such as HTML or Markdown significantly enhances their accessibility and usability. This conversion not only improves web accessibility but also boosts document interactivity, enhances searchability and indexing, and guarantees compatibility across different platforms \cite{frankston2024html}. This is particularly relevant for large-scale digital library initiatives and supports FAIR principles by making content machine-actionable \cite{Wilkinson2016}. Such documents, typically delivered in PDF format, contain complex elements including mathematical formulas, figures, headers, and tables, as well as densely layouted text, posing significant challenges for computational document processing \cite{li-etal-2020-docbank}.


Document understanding involves the automated classification and extraction of information from richly typeset documents \cite{cui2021document}. While transformer-based models \cite{LayoutLMv3} and recent end-to-end decoders \cite{blecher2023nougat, lv2023kosmos, poznanski2025olmocrunlockingtrillionstokens} can transform PDFs to Markdown, they suffer from a core inefficiency. They generate text token-by-token from scratch, leading to slow inference times and unnecessary computational overhead, especially for dense academic documents where large portions of text could be directly copied.

Despite their effectiveness, decoder-based models suffer from inefficiencies in processing speed. They generate text token-by-token from scratch, leading to slow inference times, especially for dense academic documents where large portions of text can be directly copied from PDFs. This sequential decoding approach introduces unnecessary computational overhead, making the transformation process significantly slower.



To mitigate these inefficiencies, we draw inspiration from two complementary paradigms. First, an edit-based approach, common in tasks like grammatical error correction \cite{malmi-etal-2022-text}, is more suitable than generating from scratch. Unlike sequence-to-sequence models, edit-based models identify and modify only necessary portions of a document, significantly improving inference speed. Second, to handle the generation of complex, non-copyable elements like formulas and tables, we leverage the Fill in the Middle (FIM) technique \cite{bavarian2022efficienttraininglanguagemodels}. FIM excels at inserting content while conditioning on both preceding and following context, making it ideal for our hybrid approach.

To address the inefficiencies of existing models, we introduce EditTrans. While our approach can be seen as a workflow optimization, its research novelty lies in the effective synthesis of layout-aware classification, text-editing principles, and FIM generation to directly target the redundancy bottleneck in state-of-the-art document transformers. EditTrans features a lightweight classifier, fine-tuned from a DLA model on 162,127 pages from arXiv, to create a structured edit queue that prioritizes modifications only where necessary.
In this paper, we present the following contributions:

\begin{itemize}
\item \textbf{EditTrans:} A Hybrid Editing-Generation Model that pre-identifies and queues editable text from PDFs before initiating markup generation. This approach significantly reduces redundant text generation and improves efficiency.
\item To enhance reproducibility and transparency, we release dataset-making scripts along with the arXiv document identifiers used in our experiments. Our open-source code is available online.\footnote{\url{https://github.com/Fireblossom/EditTrans}}
\item We refine PDF-to-Markup evaluation datasets, offering a fairer and more reliable benchmark for comparing tools aimed at improving scholarly document accessibility.
\end{itemize}

\section{Related Work}
\subsection{Academic Documents Transformation}
GROBID \cite{GROBID} is a machine learning library for extracting, parsing, and restructuring documents, including PDF into structured XML encoded documents.
However, it is not flexible because it converts formulas and tables into images, thus hampering subsequent accessibility.
docTR \cite{doctr2021} and DocBed \cite{DocBed} first identify the document layout and then extract text content.
Donut \cite{kim2022donut} is a Document Understanding model consisting of a visual encoder and language model decoder without obtaining texts directly from the document.
Nougat \cite{blecher2023nougat} follows Donut in implementing screenshot-to-Markdown transformation of Academic Documents. 
LOCR \cite{sun2024locr} solves the problem of Nougat's hallucination and repetition using additional location prompts.
Kosmos-2.5 \cite{lv2023kosmos} and DocOwl-1.5 \cite{hu2024docowl} implement a more generalized screenshot-to-Markdown transformation with Vision-Language methods and larger model size.
OlmOCR\cite{poznanski2025olmocrunlockingtrillionstokens} is fine-tuned from the Qwen2-VL \cite{wang2024qwen2vlenhancingvisionlanguagemodels}, which features text from PDF files and their corresponding coordinates as anchors into the language model's prompt. Closed-source services such as LlamaParse\footnote{\url{https://cloud.llamaindex.ai/parse}} and Mistral OCR \footnote{\url{https://mistral.ai/news/mistral-ocr}} also provide PDF-to-Markdown transformation.
This paper describes EditTrans, which edits Nougat, Kosmos-2.5, and OlmOCR's input sequence to speed up the transformation.

\subsection{Document Layout Analysis}
Recent Document Layout Analysis (DLA) models have become increasingly powerful thanks to the availability of large-scale document layout datasets \cite{PubLayNet, DocLayNet, li-etal-2020-docbank, FUNSD}
Computer Vision models have been able to extract layouts in screenshots of documents \cite{9412557, 9156933, 9428465}.
Language models have also been applied to recognize layouts.
LayoutLM \cite{layoutlm} and its variant VILA \cite{shen-etal-2022-vila} are transformer encoder models that analyze document layouts from the texts and their 2D coordinates.
LayoutLMv2 and 3 \cite{xu-etal-2021-layoutlmv2, LayoutLMv3} additionally attach visual features to the transformer encoder.
DiT \cite{dit} was trained autoregression from unlabeled data.
LILT \cite{wang-etal-2022-lilt} has migrated English layout pre-training to multiple languages.
ERNIE-Layout \cite{peng-etal-2022-ernie} introduces the task of reading order prediction, enhancing the accuracy of the DLA models.
In this work, our Editable Text Identification module is fine-tuned from ERNIE-Layout.

\subsection{Text Generation with Text-Editing Models}
Transformers decoder models generate outputs token-by-token from scratch, thus making them slow at inference time.
Text-editing models provide several benefits over decoder models, including faster inference speed, higher sample efficiency, as well as better control and interpretability of the outputs \cite{malmi-etal-2022-text}. LaserTagger \cite{malmi-etal-2019-encode} implements the Sentence Fusion task with three actions: \verb|KEEP|, \verb|DELETE|, and \verb|REPLACE|.
FELIX \cite{mallinson-etal-2020-felix} and EdiT5 \cite{mallinson-etal-2022-edit5} also achieve text reordering.
PIE \cite{awasthi-etal-2019-parallel}, Seq2Edits \cite{stahlberg-kumar-2020-seq2edits}, and GECToR \cite{omelianchuk-etal-2020-gector}  edit the text using the Iterative Refinement approach.
There is a blog post\footnote{\url{https://research.google/blog/grammar-checking-at-google-search-scale/}} about Google Search correcting user input using EdiT5 \cite{mallinson-etal-2022-edit5} with low-latency features. 
Additionally, recent innovations like Locate\&Edit \cite{son2024locateeditenergybasedtextediting} represent an energy-based approach to controlled text generation that can work with black-box language models by first identifying spans relevant to specific constraints and then selectively replacing them with more suitable alternatives, thereby preserving the core semantics of the original text while satisfying desired constraints.
Our work organizes PDF-sourced texts into edit queues, which mimics the behavior of text-editing models.

\subsection{Fill in the Middle}
Fill in the Middle (FIM) \cite{bavarian2022efficienttraininglanguagemodels} is a straightforward yet effective approach for training language models to fill in missing text in the middle of documents.
Their extensive evaluations demonstrated that training with this transformed data does not compromise the model’s original left-to-right generative capabilities.
FIM techniques have been applied to various natural language processing tasks, including Document Completion, Code Editing, and Writing Assistance.
Currently, large language model services for programming assistance apply the FIM paradigm, including Gemini Coder\footnote{\url{https://geminicoder.org/}}, Claude Code\footnote{\url{https://docs.anthropic.com/en/docs/agents-and-tools/claude-code/overview}}, and DeepSeek-Coder \cite{guo2024deepseekcoderlargelanguagemodel}.
The execution of editing actions in our task inherits the paradigm of FIM.

\definecolor{myblue}{RGB}{99,145,244}

\section{Methodology}
\begin{figure}[t]
    \centering
    \includegraphics[width=\textwidth]{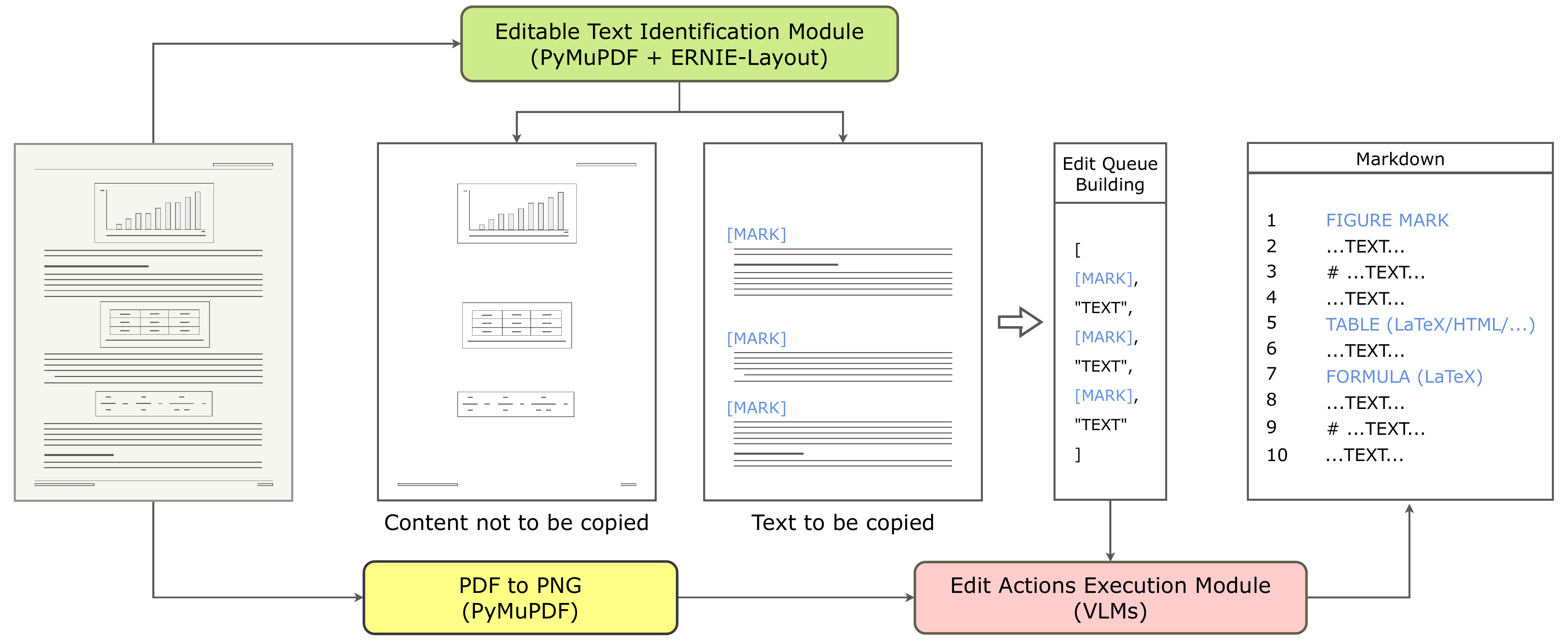}
    \cprotect\caption{The EditTrans workflow. The Editable Text Identification module detects whether the span is copyable or not. Edit Queue Building module builds an edit queue, where {\color{myblue}\verb|[MARK]|} initiates generation. Edit Actions Execution module executes edits: the {\color{myblue}blue} part is generated by a backbone VLM with the FIM paradigm.}
    \label{fig:process}
\end{figure}

EditTrans streamlines academic document transformation into three modules:
\begin{enumerate}
    \item \textbf{Editable Text Identification:} EditTrans begins by classifying spans extracted from PDF pages and identifying which portions of the spans are copyable.
    \item \textbf{Edit Queue Building:} EditTrans then organizes the classified spans into an edit action queue and delineates a stop criterion for each edit needed.
    \item \textbf{Edit Actions Execution:} For each span requiring editing, EditTrans utilizes the pre-trained PDF-to-Markdown models to execute the necessary edits with a FIM paradigm.
\end{enumerate}

Figure~\ref{fig:process} briefly demonstrates how we can copy text from a PDF and save end-to-end PDF-to-Markdown model's inference steps and reduce latency.
Figure~\ref{fig:steps} shows an example of EditTrans processing inline formulas.

\subsection{Editable Text Identification}
\label{sec:tei}
Inspired by DLA-related work, we assume that whether the text is copyable or not is highly correlated with its layout information.
Specifically, we suggest the following:
\begin{enumerate}
    \item Dense plain text found in paragraphs should be preserved in its entirety.
    \item Page elements such as mathematical formulas, tables, and titles should be modified to align with Markdown formatting standards.
    \item Elements that do not convey relevant content, including page headers, footers, and page numbers, should be excluded from the final document.
\end{enumerate}

Following VILA \cite{shen-etal-2022-vila}, we assume that text editability is homogeneous at the span level.
We use PyMuPDF\footnote{\url{https://github.com/pymupdf/PyMuPDF}} to extract span-level text and bounding boxes from the PDF.
Subsequently, we use PyMuPDF4LLM\footnote{\url{https://pymupdf.readthedocs.io/en/latest/pymupdf4llm/}}, a heuristic to detect the reading order of text spans.

We fine-tune the ERNIE layout model for token classification using LoRA \cite{hu2022lora}, omitting global 1D position embeddings to prevent potential biases in layout judgment~\cite{tu-etal-2023-layoutmask}.
This module simplifies the target from the broader range of layout categories typically used in DLA models (e.g., up to 10 categories in DocBank~\cite{li-etal-2020-docbank}) to 3 categories.
We altered ERNIE-Layout's classification head to predict labels as \verb|KEEP|, \verb|DELETE|, or \verb|INSERT_LEFT|.
\begin{itemize}
    \item \verb|KEEP| indicates that the span should be included in the Markdown output without editing;
    \item \verb|DELETE| indicates that the span should be deleted;
    \item \verb|INSERT_LEFT| indicates that a trigger for PDF-to-Markdown model generation should be inserted before this span. 
\end{itemize}

This approach is notably different from already existing text-editing models. 
Our editing logic in this paper is to delete text that should not be copied and then insert tokens that should be generated by the decoder, e.g., a mathematical formula is deleted from the PDF, and an equivalent expression is generated before the next span of editable text.

As we have fine-tuned the ERNIE layout for token classification, and each span may contain more than one token, a voting classifier \cite{6065432} is applied to decide the prediction of the spans. 
Details of the fine-tuning hyperparameters are documented in Appendix~\ref{sec:hyp}.

\subsection{Edit Queue Building}
Once we have the span-level edit annotation, we can turn it into an edit queue $Q$ that prompts the pre-trained PDF-to-Markdown model for which portion of the text to edit.
Each edit queue $Q$ starts with an edit trigger, followed by a sequential processing of each span.
We iterate through each span in the edit annotation.

If the next span is predicted to label \verb|KEEP|, we add span text to $Q$. 
Note that if the length of the text characters in this span is very small, we do not add it and instead expect the PDF-to-Markdown model to generate it because a text makes it difficult to match where the PDF-to-Markdown model should stop generating.
We then match the first words in the text of this span to sign the PDF-to-Markdown model stop generation and start copy.

If the next span is predicted as \verb|DELETE|, we do not add anything to $Q$.

If the next span is \verb|INSERT_LFET|, we will add an edit trigger first and then add this span's text sequence to $Q$. 
Similarly to the \verb|KEEP| span, we will match the first words as an edit stop sign to the PDF-to-Markdown model.
Note that if the length of the text characters in this span is very small (e.g. fewer than 5 characters), we do not add it and instead expect the PDF-to-Markdown model to generate it. 
This heuristic prevents inefficient, fragmented generation where the model would be frequently interrupted to copy very short spans.

At the end, we will add an extra edit trigger to allow the PDF-to-Markdown model to generate end-of-sentence tokens.

\subsection{Edit Actions Execution}
\label{sec:eae}
Algorithm~\ref{alg:edit-actions} briefly demonstrates the execution of the edit action.
In this step, we initialize an empty tokens sequence $S$ and traverse the edit queue $Q$.

If the next element in $Q$ is an edit trigger, a screenshot of the page and $S$ are fed into the PDF-to-Markdown model.
The PDF-to-Markdown model’s generation phase involves Fill Markdown In the Middle task, which predicts the next token in the sequence until a stopping criterion is met.
The output of the PDF-to-Markdown model is added to $S$.

If the next element in the queue is a to-be-copied span, we tokenize the text of this span and add it to the end of $S$.

\begin{algorithm}[t]
\caption{Edit actions execution process, the example end-to-end PDF-to-Markdown model in this pseudo code is Nougat.}
\label{alg:edit-actions}
\KwIn{Edit actions queue $Q$, PDF document $D$, Nougat model $\mathcal{M}$}
\KwOut{Generated Markdown sequence $S$}

Initialize empty token sequence $S$\;
Extract screenshot of the page from $D$\;
\While{$Q$ is not empty}{
    $q \leftarrow$ dequeue($Q$)\;
    
    \If{$q$ is an edit trigger}{
        Feed screenshot and $S$ into $\mathcal{M}$\;
        \While{$\mathcal{M}$ has not reached stopping criteria}{
            Generate next token and append to $S$\;
        }
    }\
    \ElseIf{$q$ is a to-be-copied span}{
        Tokenize text from $q$ and append to $S$\;
    }
}
Detokenize $S$ into Markdown format\;
\Return $S$\;
\end{algorithm}

\paragraph{Fill Markdown In the Middle.}
In this module, the pre-trained PDF model is fed with the current in-progress token sequence and begins to autoregressively decode through the decoding until it departs the stopping criteria.
We set the stopping criteria to be the first $n$ tokens of the next-to-be-copied span or end-of-sentence tokens.

\paragraph{Batch Synchronization.}
For batch inference, since the lengths of the next-to-be-copied spans in the edit queue $Q$ are not equal, the lengths of the sequences 
in $S$ are not equal after executing each edit action.
We perform left padding for each sequence within the batch to ensure that the operations within the batch are synchronized.

\paragraph{Edit Actions Skip.}
\label{sec:eas}
In practice, we have found some deviations in the Edit Action Queue. 
Specifically, the recognition of Editable Text Identification (ETI) may not be reliable enough, and as a result, it may produce an incorrect edit action queue.
In order to minimize the impact of ETI errors, we added a function to skip the current action in the Edit Actions Execution module.
If the last $n' > n$ tokens currently generated coincide with the next-to-be-copied span in the edit action queue, we pop the next action from the edit action queue $Q$ and proceed to execute it. 

The values for $n$ and $n'$ (empirically set to 3 and 5, respectively) represent a trade-off. Smaller values risk premature termination of generation, while larger values might cause the model to overlook necessary small edits.

In a nutshell, we copy the simple text from the PDF and leave PDF-to-Markdown models in charge of generating the complex parts, such as formulas and tables.
Finally, $S$ is outputted and de-tokenized into the Markdown format.

EditTrans' hyperparameters can be found in Section~\ref{sec:hyperparam}. 

\section{Experiments}
\subsection{PDF-to-Markdown models}
\label{sec:models}
There are currently three open-source weighted PDF-to-Markdown models.
We used them as backbones to test the validity of EditTrans.

\begin{enumerate}
    \item Nougat \cite{blecher2023nougat}, we use \verb|nougat-base|\footnote{\url{https://huggingface.co/facebook/nougat-base}} which has 349M parameters.
    \item Kosmos-2.5\footnote{\url{https://huggingface.co/microsoft/kosmos-2.5}} \cite{lv2023kosmos}, with 1.37B parameters.
    \item OlmOCR \cite{poznanski2025olmocrunlockingtrillionstokens}, we use \verb|olmOCR-7B-0225-preview|\footnote{\url{https://huggingface.co/allenai/olmOCR-7B-0225-preview}} with 8.29B parameters.
\end{enumerate}
EditTrans is expected to produce the same output as Nougat, Kosmos-2.5, and OlmOCR (i.e., generate Markdown).
EditTrans requires a PDF page as input, while Nougat and Kosmos-2.5 require a screenshot of the PDF page, and OlmOCR additionally requires text anchors in PDF files to build the prompt. 

Prompt Lookup Decoding (PLD) \cite{saxena2023prompt} is a technique that speeds up LLM generation by identifying and reusing sequences of tokens that appear in the input prompt.
Since OlmOCR uses text anchors in the file as part of the prompt, we use it as a comparison. Nougat and Kosmos-2.5 do not apply to PLD because they don't have text in the prompt.

\subsection{Datasets Building}
As there is no existing dataset released as full PDFs at this time,
we downloaded the \LaTeX{} source code bundles for the July and August 2023 papers from arXiv.
Then we use a framework \cite{duan-etal-2023-latex} that compiles \LaTeX{} to PDF, plus annotates for semantic labels, reading order, and \LaTeX{} code corresponding to mathematical formulas and tables for each element on a page.
A part of the downloaded source code of the papers was not annotated successfully because it was written in a way that the framework could not parse.
A total number of 14,320 papers were annotated, yielding 180,146 pages.

Spans are extracted from these pages and are labeled as either \verb|KEEP|, \verb|DELETE|, or \verb|INSERT_LEFT|, based on the results of the semantic annotation of the previous step.
We mark the captions of figures and tables as \verb|DELETE| because they are reordered.

Pages that are empty or challenging to read, such as those containing full-page images, long tables, or bibliographies, are excluded from the dataset.
Finally, a dataset was assembled consisting of 180.146 pages, each annotated with span-level text copyable labels and their corresponding bounding boxes.
Our dataset is rich in layout diversity, including single- and multi-column formats, complex tables, numerous mathematical equations, and embedded figures, allowing robust evaluation beyond simple layouts.
We randomly split the training set size to 162,127 and the test set to 18,019.
The vast majority of the pages in this dataset are in English.

We then attached a Markdown target for each page, which emulates the PDF-to-Markdown model's style of inserting mathematics formulas and tables as \LaTeX{} code.
\LaTeX{} is quite flexible because it allows user-defined macros.
Therefore, we normalize the formula and table \LaTeX{} codes with LaTeXML\footnote{\url{https://math.nist.gov/~BMiller/LaTeXML/}}. 

The method in this paper extracts text spans from PDFs, which requires access to the full text of academic papers.
As arXiv does not grant permission to repost the full-text\footnote{\url{https://info.arxiv.org/help/license/reuse.html\#full\_text}}, we publish the scripts for creating the datasets plus the dataset's arXiv numbers to provide reproducibility.

\subsection{Model-specific PDF-to-Markdown Test Dataset Building}
In previous work, since neither Nougat nor Kosmos-2.5 released their datasets, transformation quality was evaluated on their private datasets. 
OlmOCR, on the other hand, implements human evaluations.
These are difficult to reproduce.

We randomly selected 128 pages from the test set for the PDF-to-Markdown transformation speed and quality test. 
In addition, we also randomly selected 128 pages each from papers categorized in Economics and Quantum Physics, respectively, to test the performance of EditTrans on different domains.

In practice, we found that the output formats of the three models are not the same, depending on their training strategies.
\begin{enumerate}
    \item Nougat transforms all the mathematical formulas in the PDF page to \LaTeX{} code. Nougat then takes the table body, as well as the caption of the graph and table, and places it at the end of the Markdown of the page.
    \item Kosmos-2.5 converts the table in the PDF page to HTML format\footnote{This observed behavior is different from that described in their paper \cite{lv2023kosmos}.}, and it places the table and its caption at the beginning of the page. Kosmos-2.5 then places the caption of the figure at the tail end of the page.
    \item OlmOCR will arrange tables and figures in the order in which they appear on the page and convert the tables with Markdown table syntax.
\end{enumerate}
However, these differences do not lead to worse readability of the transformation results.
Based on the fact that the PDF-to-Markdown transformation task is relatively flexible, we matched the behaviour of each model and assembled appropriate standard transformation results for each of them.

This provides fair and reproducible criteria for transformation quality.
The scripts for making the test dataset are open-sourced on GitHub in order to provide a fair evaluation for future PDF-to-Markdown models as well.

\subsection{Evaluation Metrics}
We performed timings on each module to evaluate the transformation latency reduced by EditTrans.
Following Nougat, we use Edit Distance \cite{Levenshtein_SPD66}, 
BLEU \cite{papineni-etal-2002-bleu}, 
METEOR \cite{banerjee-lavie-2005-meteor}, precision, recall, and F-1 to evaluate transformation quality. 

\textbf{Edit Distance} calculates the minimum number of character edits (insertions, deletions, substitutions) required to convert one text into another, expressed as a ratio of total characters to standardize comparisons.

\textbf{BLEU} is a widely-used evaluation metric in machine translation tasks, BLEU assesses textual similarity by counting overlapping sequences of words (n-grams) between the predicted and reference texts.

\textbf{METEOR} is a translation evaluation metric emphasizing recall. It measures semantic alignment between the predicted text and reference text, considering synonym matches and paraphrased content.

\textbf{Precision, recall, and F-1} provide a balanced measure of accuracy and completeness in the generated text predictions.

\section{Results}
\subsection{Editable Text Identification Model Accuracy}
We evaluate the classifier's performance using token-level F1 scores with micro-averaging. The achieved token-level F1 score is \textbf{0.963}. The classifier demonstrates efficient inference, requiring approximately 0.03 seconds per page.

\begin{table}[]
\caption{Generation steps and inference delays for three models on three test datasets. w/o ET represents models directly reading PDF files and generating their Markdown code. w/ ET represents models generated with the help of EditTrans. OlmOCR additionally compares the efficiency of PLD.}
\label{tab:latency}
\centering
\begin{adjustbox}{width=1\textwidth}
\begin{tabular}{l|c|cccc}
\hline
 &  & \multicolumn{4}{c}{Metrics} \\ \hline
Model & Test Dataset & \multicolumn{1}{c|}{Generation Steps} & \multicolumn{1}{c|}{Saving Steps} & \multicolumn{1}{c|}{Latency (s)} & Reduced latency \\ \hline
Nougat w/o ET &  & 926.82 & - & 6.37 & - \\
\cellcolor[HTML]{EFEFEF}Nougat w/ ET &  & \cellcolor[HTML]{EFEFEF}656.28 & \cellcolor[HTML]{EFEFEF}29.2\% & \cellcolor[HTML]{EFEFEF}4.47 & \cellcolor[HTML]{EFEFEF}29.8\% \\
Kosmos-2.5 w/o ET &  & 879.28 & - & 11.62 & - \\
\cellcolor[HTML]{EFEFEF}Kosmos-2.5 w/ ET &  & \cellcolor[HTML]{EFEFEF}555.88 & \cellcolor[HTML]{EFEFEF}36.8\% & \cellcolor[HTML]{EFEFEF}7.24 & \cellcolor[HTML]{EFEFEF}37.7\% \\
OlmOCR w/o ET &  & 1148.47 & - & 34.47 & - \\
\cellcolor[HTML]{EFEFEF}OlmOCR w/ ET & \multirow{-6}{*}{arXiv} & \cellcolor[HTML]{EFEFEF}766.69 & \cellcolor[HTML]{EFEFEF}33.2\% & \cellcolor[HTML]{EFEFEF}22.91 & \cellcolor[HTML]{EFEFEF}33.5\% \\ 
\hline
Nougat w/o ET &  & 817.62 & - & 5.63 & - \\
\cellcolor[HTML]{EFEFEF}Nougat w/ ET &  & \cellcolor[HTML]{EFEFEF}562.17 & \cellcolor[HTML]{EFEFEF}31.2\% & \cellcolor[HTML]{EFEFEF}3.84 & \cellcolor[HTML]{EFEFEF}31.7\% \\
Kosmos-2.5 w/o ET &  & 786.66 & - & 10.38 & - \\
\cellcolor[HTML]{EFEFEF}Kosmos-2.5 w/ ET &  & \cellcolor[HTML]{EFEFEF}446.27 & \cellcolor[HTML]{EFEFEF}43.3\% & \cellcolor[HTML]{EFEFEF}5.81 & \cellcolor[HTML]{EFEFEF}44.0\% \\
OlmOCR w/o ET &  & 870.08 & - & 26.42 & - \\
\cellcolor[HTML]{EFEFEF}OlmOCR w/ ET & \multirow{-6}{*}{Economics} & \cellcolor[HTML]{EFEFEF}561.82 & \cellcolor[HTML]{EFEFEF}35.4\% & \cellcolor[HTML]
{EFEFEF}17.07 & \cellcolor[HTML]{EFEFEF}35.3\% \\ 
\hline
Nougat w/o ET &  & 967.80 & - & 6.59 & - \\
\cellcolor[HTML]{EFEFEF}Nougat w/ ET &  & \cellcolor[HTML]{EFEFEF}607.63 & \cellcolor[HTML]{EFEFEF}37.2\% & \cellcolor[HTML]{EFEFEF}4.12 & \cellcolor[HTML]{EFEFEF}37.4\% \\
Kosmos-2.5 w/o ET &  & 911.75 & - & 11.99 & - \\
\cellcolor[HTML]{EFEFEF}Kosmos-2.5 w/ ET &  & \cellcolor[HTML]{EFEFEF}511.97 & \cellcolor[HTML]{EFEFEF}43.8\% & \cellcolor[HTML]{EFEFEF}6.66 & \cellcolor[HTML]{EFEFEF}44.5\% \\
OlmOCR w/o ET &  & 1174.38 & - & 35.73 & - \\
\cellcolor[HTML]{EFEFEF}OlmOCR w/ ET & \multirow{-6}{*}{\begin{tabular}[c]{@{}c@{}}Quantum\\ Physics\end{tabular}} & \cellcolor[HTML]{EFEFEF}738.45 & \cellcolor[HTML]{EFEFEF}37.1\% & \cellcolor[HTML]{EFEFEF}22.32 & \cellcolor[HTML]{EFEFEF}37.5\% \\
\hline
\end{tabular}
\end{adjustbox}
\end{table}

\begin{table}[]
\caption{PDF-to-Markdown transformation quality of the three models on the three test datasets. w/o ET represents the model directly reading and generating the screenshots of the PDF file. w/ ET represents the model generated with the help of EditTrans.}
\label{tab:quality}
\centering
\begin{adjustbox}{width=1\textwidth}
\begin{tabular}{l|c|cccccc}
\hline
 &  & \multicolumn{6}{c}{Metrics} \\ \hline
Model & Test Dataset & \multicolumn{1}{c|}{Edit Dist $\downarrow$} & \multicolumn{1}{c|}{BLEU $\uparrow$} & \multicolumn{1}{c|}{METEOR $\uparrow$} & \multicolumn{1}{c|}{Precision $\uparrow$} & \multicolumn{1}{c|}{Recall $\uparrow$} & F1 $\uparrow$ \\ \hline
Nougat w/o ET &  & 0.343 & 0.547 & 0.616 & 0.822 & 0.675 & \textbf{0.720} \\
\cellcolor[HTML]{EFEFEF}Nougat w/ ET &  & \cellcolor[HTML]{EFEFEF}\textbf{0.335} & \cellcolor[HTML]{EFEFEF}0.542 & \cellcolor[HTML]{EFEFEF}0.619 & \cellcolor[HTML]{EFEFEF}0.801 & \cellcolor[HTML]{EFEFEF}0.676 & \cellcolor[HTML]{EFEFEF}0.715 \\
Kosmos-2.5 w/o ET &  & \textbf{0.384} & 0.506 & 0.590 & 0.724 & 0.634 & 0.656 \\
\cellcolor[HTML]{EFEFEF}Kosmos-2.5 w/ ET &  & \cellcolor[HTML]{EFEFEF}0.387 & \cellcolor[HTML]{EFEFEF}0.516 & \cellcolor[HTML]{EFEFEF}0.600 & \cellcolor[HTML]{EFEFEF}0.737 & \cellcolor[HTML]{EFEFEF}0.644 & \cellcolor[HTML]{EFEFEF}\textbf{0.666} \\
OlmOCR w/o ET &  & 0.311 & 0.541 & 0.676 & 0.671 & 0.679 & 0.667 \\
\cellcolor[HTML]{EFEFEF}OlmOCR w/ ET & \multirow{-6}{*}{arXiv} & \cellcolor[HTML]{EFEFEF}\textbf{0.294} & \cellcolor[HTML]{EFEFEF}0.532 & \cellcolor[HTML]{EFEFEF}0.683 & \cellcolor[HTML]{EFEFEF}0.698 & \cellcolor[HTML]{EFEFEF}0.699 & \cellcolor[HTML]{EFEFEF}\textbf{0.688} \\ \hline
Nougat w/o ET &  & 0.283 & 0.645 & 0.710 & 0.838 & 0.759 & \textbf{0.778} \\
\cellcolor[HTML]{EFEFEF}Nougat w/ ET &  & \cellcolor[HTML]{EFEFEF}\textbf{0.278} & \cellcolor[HTML]{EFEFEF}0.625 & \cellcolor[HTML]{EFEFEF}0.701 & \cellcolor[HTML]{EFEFEF}0.819 & \cellcolor[HTML]{EFEFEF}0.746 & \cellcolor[HTML]{EFEFEF}0.762 \\
Kosmos-2.5 w/o ET &  & 0.320 & 0.590 & 0.699 & 0.751 & 0.738 & 0.726 \\
\cellcolor[HTML]{EFEFEF}Kosmos-2.5 w/ ET &  & \cellcolor[HTML]{EFEFEF}\textbf{0.312} & \cellcolor[HTML]{EFEFEF}0.604 & \cellcolor[HTML]{EFEFEF}0.706 & \cellcolor[HTML]{EFEFEF}0.768 & \cellcolor[HTML]{EFEFEF}0.742 & \cellcolor[HTML]{EFEFEF}\textbf{0.739} \\
OlmOCR w/o ET &  & 0.281 & 0.594 & 0.742 & 0.707 & 0.737 & 0.712 \\
\cellcolor[HTML]{EFEFEF}OlmOCR w/ ET & \multirow{-6}{*}{Economics} & \cellcolor[HTML]{EFEFEF}\textbf{0.277} & \cellcolor[HTML]{EFEFEF}0.584 & \cellcolor[HTML]{EFEFEF}0.734 & \cellcolor[HTML]{EFEFEF}0.725 & \cellcolor[HTML]{EFEFEF}0.760 & \cellcolor[HTML]{EFEFEF}\textbf{0.732} \\ \hline
Nougat w/o ET &  & 0.356 & 0.528 & 0.580 & 0.838 & 0.667 & 0.729 \\
\cellcolor[HTML]{EFEFEF}Nougat w/ ET &  & \cellcolor[HTML]{EFEFEF}\textbf{0.349} & \cellcolor[HTML]{EFEFEF}0.550 & \cellcolor[HTML]{EFEFEF}0.601 & \cellcolor[HTML]{EFEFEF}0.863 & \cellcolor[HTML]{EFEFEF}0.688 & \cellcolor[HTML]{EFEFEF}\textbf{0.752} \\
Kosmos-2.5 w/o ET &  & 0.384 & 0.528 & 0.600 & 0.778 & 0.660 & 0.696 \\
\cellcolor[HTML]{EFEFEF}Kosmos-2.5 w/ ET &  & \cellcolor[HTML]{EFEFEF}\textbf{0.374} & \cellcolor[HTML]{EFEFEF}0.545 & \cellcolor[HTML]{EFEFEF}0.616 & \cellcolor[HTML]{EFEFEF}0.794 & \cellcolor[HTML]{EFEFEF}0.671 & \cellcolor[HTML]{EFEFEF}\textbf{0.713} \\
OlmOCR w/o ET &  & 0.294 & 0.573 & 0.699 & 0.715 & 0.707 & 0.704 \\
\cellcolor[HTML]{EFEFEF}OlmOCR w/ ET & \multirow{-6}{*}{\begin{tabular}[c]{@{}c@{}}Quantum\\ Physics\end{tabular}} & \cellcolor[HTML]{EFEFEF}\textbf{0.257} & \cellcolor[HTML]{EFEFEF}0.571 & \cellcolor[HTML]{EFEFEF}0.712 & \cellcolor[HTML]{EFEFEF}0.735 & \cellcolor[HTML]{EFEFEF}0.712 & \cellcolor[HTML]{EFEFEF}\textbf{0.708} \\ \hline
\end{tabular}
\end{adjustbox}
\end{table}

\subsection{PDF-to-Markdown Transformation Latency and Quality}
Table~\ref{tab:latency} shows that EditTrans saves inference steps and thus reduces latency.
On average, Kosmos-2.5 gains the greatest advantage from EditTrans, reducing generation steps by up to 43.8\% on the Quantum Physics dataset and by at least 36.8\% on the arXiv dataset. Conversely, Nougat benefits the least, with step reductions as low as 29.2\% on the arXiv dataset. Overall, the reduction in latency closely aligns with the proportion of inference steps saved. Also, we observe that PLD reduces the latency of OlmOCR, but not as efficiently as EditTrans.

Table~\ref{tab:quality} illustrates that EditTrans generally improves or only minimally affects the quality of PDF-to-Markdown transformation.
In terms of Edit Distance, EditTrans typically reduces errors across datasets; however, a minor increase of 0.003 occurs for Kosmos-2.5 on the arXiv dataset. Translation-based metrics (BLEU and METEOR) exhibit negligible differences, with variations remaining below 0.01, indicating a tiny substantial impact from EditTrans. 
The F1 scores mostly show improvements when EditTrans is applied, except in the cases of Nougat on the arXiv and Economics datasets with 0.005. 
Overall, these results demonstrate that EditTrans consistently maintains or marginally enhances transformation quality without any significant negative effects.
We do not add the PLD comparison to the table as it does not change the generated texts.

\begin{figure}
    \centering
    \begin{adjustbox}{center}
        \includegraphics[width=\linewidth]{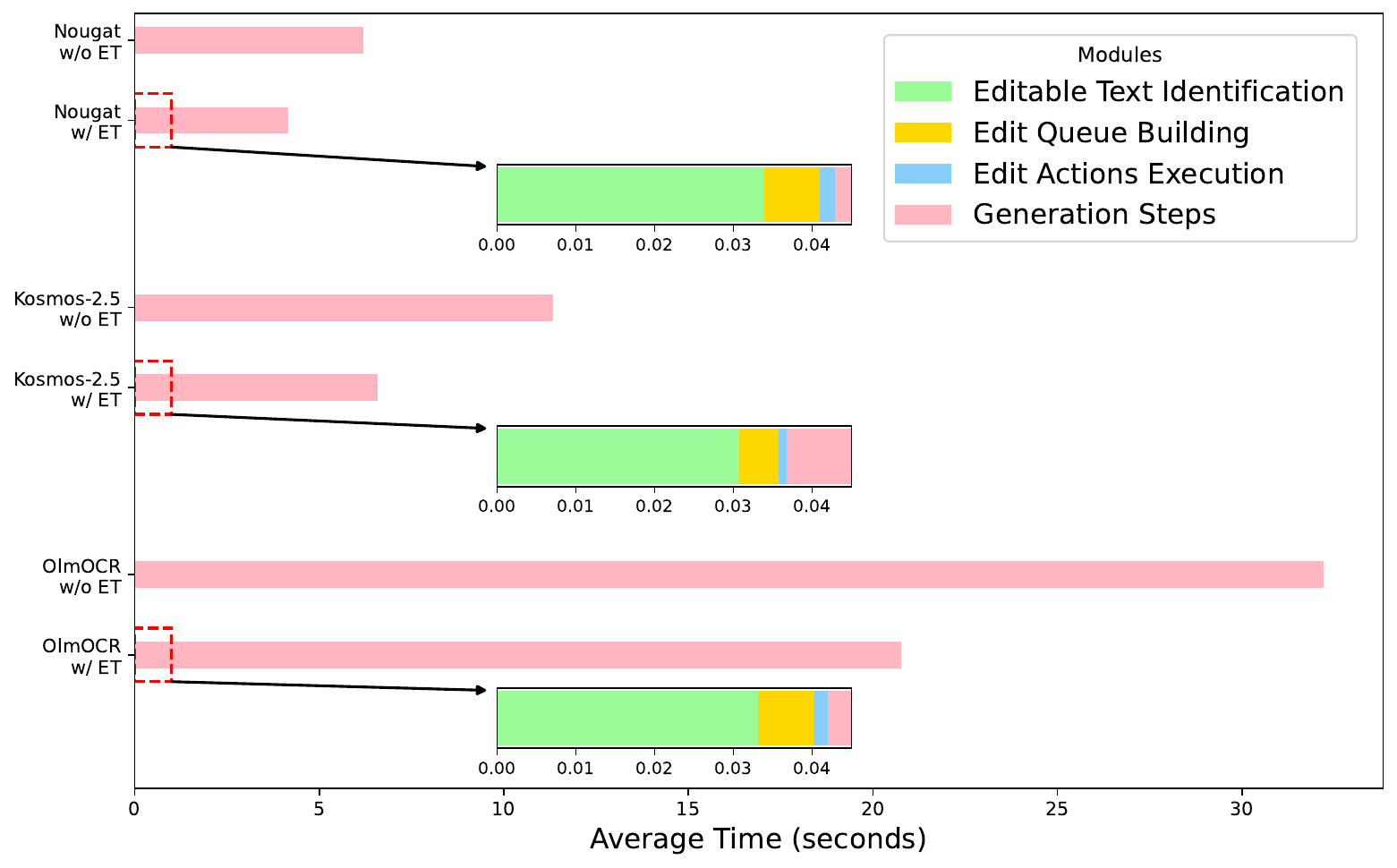}
    \end{adjustbox}
    \caption{Execution time overhead for each module. where ET stands for EditTrans. Generation Step in w/ ET represents the Fill Markdown In the Middle step described in Section~\ref{sec:eae}, and in w/o ET represents the inference step of the model. As can be seen from the figure, the module of EditTrans adds an extra portion of latency ($\sim$0.04 seconds), but it is insignificant compared to the time it saves in the Generation Step.}
    \label{fig:latency}
\end{figure}
\begin{figure}
    \centering
    \includegraphics[width=0.9\linewidth]{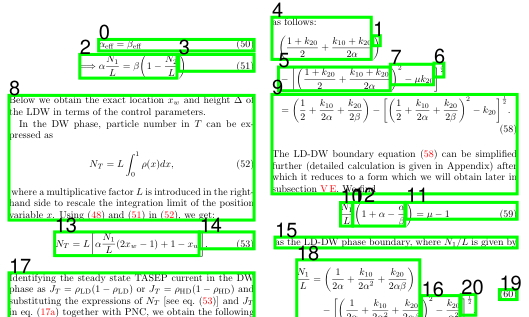}
    \caption{Edge case example encountered during the construction of the editing action queue. The green boxes represent text spans, and the numbers indicate the reading order predicted by PyMuPDF4LLM. In this example, the text columns on the left should have been read from top to bottom, followed by the columns on the right. But their reading order is disrupted. Causing EditTrans becomes ineffective in this scenario.}
    \label{fig:fail}
\end{figure}

\subsection{Latency of Each Module in EditTran}
We analyze the execution time overhead introduced by each module as an ablation study.
As illustrated in Figure~\ref{fig:latency}, integrating EditTrans incurs a slight latency increase. 
However, this overhead is negligible when compared to the substantial time savings achieved during the generation step. 
Importantly, this additional latency remains stable and does not escalate with larger backbone model sizes.

\subsection{Error Analysis}
\paragraph{Mitigating ETI Error Propagation.}
A core challenge is that errors from the upstream ETI module could propagate. We designed the \textit{Edit Actions Skip} mechanism (\ref{sec:eas}) specifically to address this. This feature acts as a self-correction step. For example, if the ETI module incorrectly labels a span for generation, but the backbone model generates the content of the subsequent \verb|KEEP| span anyway, our skip mechanism detects this overlap and prevents the redundant copy action. This allows the pipeline to gracefully recover from ETI errors, enhancing its robustness.

\paragraph{Garbage in, garbage out.}
As Figure~\ref{fig:fail} shows, PyMuPDF4LLM, while generally reliable, can struggle with extremely complex PDF layouts. This example is a very rare edge case ($<$0.03\% of our dataset pages), where PyMuPDF4LLM's spatial heuristics for span extraction misinterpret the true reading order. Our current adoption of PyMuPDF4LLM was a deliberate choice, carefully balancing its generally robust accuracy and high processing efficiency across the vast majority of documents against the challenge posed by such infrequent, complex cases. Consequently, when these rare reading order errors do occur, the edit action queue generated by EditTrans becomes invalid, leading to no reduction in generation steps for those specific pages. We notice that the OlmOCR method of extracting text anchors seems to extract text in the correct reading order. We will turn to this method in future studies.

\paragraph{A small vocabulary is better.}
During quality assessments, one frequent error observed with Kosmos-2.5 was the incorrect representation of \LaTeX{} codes for inline mathematical formulas. For instance, the symbol $\pm$ should be encoded as \verb|\pm| according to \LaTeX{} standards, but Kosmos-2.5 instead outputs the Unicode character \verb|\u00b1|. This issue likely arises because Kosmos-2.5 uses an extensive vocabulary, and its pre-training dataset may not have effectively filtered out such specific encoding errors compared to models like Nougat.

\paragraph{1024 tokens are not enough.}
The maximum generation length for Nougat and Kosmos-2.5 is set to 1024, while OlmOCR is set to 2048. 
This is due to the different maximum lengths of their models and the limitation of GPU RAM.
Table~\ref{tab:length} illustrates that the maximum sequence length of 1024 tokens is generally inadequate for fully converting a single page of scientific documents. Up to 80.5\% of pages encountered truncation due to this limit. Although EditTrans occasionally allows individual outputs to slightly exceed this limit through copy-span operations, it does not fundamentally solve the truncation problem. Even OlmOCR’s increased token limit of 2048 remains insufficient in 10.2\% of cases on the arXiv dataset. Consequently, sequence length limitations continue to pose significant challenges in PDF-to-Markdown transformations. Moreover, we observed that tables, followed by complex formulas, consume substantial token resources. Developing a language-model-friendly standard specifically for tables could potentially alleviate this issue.

\begin{table}[]
\centering
\caption{Pre-trained PDF-to-Markdown models sometimes produce outputs that reach the maximum sequence length limit. When this occurs, texts at the end of the page is truncated, potentially leading to data loss and incomplete transformations.}
\label{tab:length}
\begin{tabular}{l|ccc}
\hline
 & \multicolumn{3}{c}{Test Dataset} \\ \hline
Model & \multicolumn{1}{c|}{arXiv} & \multicolumn{1}{c|}{Economics} & Quantum Physics \\ \hline
Nougat & 67.2\% & 43.8\% & 80.5\% \\
Kosmos-2.5 & 54.7\% & 37.5\% & 62.5\% \\
OlmOCR & 10.2\% & 1.6\% & 3.9\% \\ \hline
\end{tabular}
\end{table}

\paragraph{Specialized models are more stable.}
During our quality assessment of OlmOCR's outputs, we identified several consistency issues. (1) OlmOCR occasionally generates redundant or unclear image captions; for example, \verb|![Figure 5: ...](image)| is Markdown-compliant but contains semantically meaningless references. (2) OlmOCR inconsistently uses both \verb|$...$| and \verb|\(...\)| notations for inline mathematical expressions. While both formats are valid in \LaTeX, mixing them complicates quality control. These inconsistencies primarily result from OlmOCR's fine-tuning based on Qwen2-VL, which is a model pre-trained on highly diverse common sense datasets, introducing unwanted variability and potentially affecting performance on specialized tasks.

\section{Future Work}
We observed that certain documents cannot be fully transformed by existing PDF-to-Markdown models due to issues such as hallucinations and repetitive text. These problems persist even with EditTrans, as it does not directly influence the backbone model's generation process. LOCR \cite{sun2024locr} addresses these issues effectively by enhancing Nougat's output using visual positional guidance, significantly reducing occurrences of hallucination and repetition. Given LOCR's complementary nature to Nougat, it should integrate smoothly with EditTrans. We are closely monitoring LOCR's progress and plan to combine it with EditTrans upon its official release.

Another significant challenge is that current PDF-to-Markdown models typically exclude figures from the transformation, whereas ERNIE Layout has proven capable of accurate figure extraction. We plan to explore methods for embedding figures into Markdown outputs effectively. One promising approach is inspired by Mistral OCR, which encodes images as base64 strings directly within Markdown documents.

Additionally, given the success of FIM methods in code generation, Large Language Diffusion Models \cite{nie2025largelanguagediffusionmodels} present an intriguing area for future research. This approach could be particularly suitable because PDF-to-Markdown tasks primarily leverage visual and textual information rather than relying solely on linguistic sequences.

User-centered assessments focusing on the readability and scholarly usefulness of the transformed documents would provide valuable insights into the practical impact of our method.
Tree-based edit distance \cite{xu2020table} is also an evaluation metric that can be extended to Markdown.




\section{Conclusion}
In this paper, we try to address the critical inefficiency of end-to-end models for the task of transforming academic PDFs into structured Markdown. We introduced EditTrans, a hybrid editing-generation framework that reframes the problem by decoupling simple text copying from complex, generative tasks. By leveraging a lightweight, layout-aware classifier to build an edit queue and then directing powerful, off-the-shelf models to perform targeted Fill-in-the-Middle generation, EditTrans achieves significant efficiency gains. Our comprehensive experiments demonstrated a latency reduction of up to 44.5\% and a saving of over 43\% in decoding steps across multiple state-of-the-art backbone models, all while maintaining or slightly improving the quality and fidelity of the final output.

The significance of this work extends beyond mere acceleration. EditTrans offers a practical pathway to scalable document processing for the digital library and archival communities. By drastically reducing the computational cost of conversion without sacrificing quality, it makes the high-fidelity digitization of vast scholarly collections—a cornerstone of modern research infrastructure—both economically and temporally feasible. This directly supports the FAIR principles by transforming static, difficult-to-parse PDFs into accessible, searchable, and machine-actionable content, thereby enhancing their value for researchers, preservationists, and the broader academic community.

While we acknowledge the limitations of our current implementation, such as its dependency on upstream reading order tools and its focus on textual content, the EditTrans framework establishes a robust and modular foundation. Its design inherently allows for future enhancements, including the integration of more advanced layout analysis and figure processing modules. Ultimately, EditTrans represents a practical and impactful step towards creating more intelligent, efficient, and scalable workflows for preserving and disseminating scholarly knowledge in the digital age.

\section*{Acknowledgments}
This work was conducted within the research project InsightsNet (\url{insightsnet.org}), which is funded by the Federal Ministry of Education and Research (BMBF) under grant no. 01UG2130A.
We gratefully acknowledge support from Dr. Wolfgang Stille and the hessian.AI Service Center (funded by the Federal Ministry of Education and Research, BMBF, grant no. 01IS22091) and the hessian.AI Innovation Lab (funded by the Hessian Ministry for Digital Strategy and Innovation, grant no. S-DIW04/0013/003).
We want to thank Dr. Sabine Bartsch for supervising this research.
We also thank the anonymous reviewers for their insightful comments and suggestions, which greatly improved the manuscript.

\bibliographystyle{splncs04}
\bibliography{anthology,custom} 

\appendix

\section{Example}
\begin{figure}[]
    \centering
    \includegraphics[width=\textwidth]{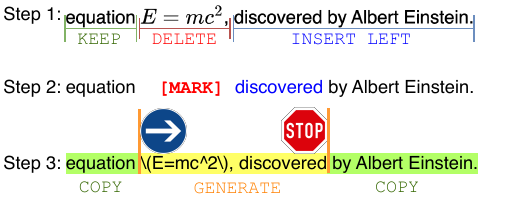}
    \cprotect\caption{Example of Edit Trans processing inline formulas. Step 1 detects whether the span is copyable or not. Step 2 builds an edit queue, where {\color{red}\verb|[MARK]|} initiates generation and a {\color{blue} \verb|blue word|} serves as a stop signal. Step 3 executes edits: the \colorbox{green}{green} part is copied, while the \colorbox{yellow}{yellow} part is generated by a backbone model with the Fill in the Middle paradigm.}
    \label{fig:steps}
\end{figure}

\section{Details for Fine-tuning ERNIE Layout}
\label{sec:hyp}
\begin{itemize}
    \item[*] Base Model: \verb|ERNIE-Layout-PyTorch|\footnote{\url{https://huggingface.co/Norm/ERNIE-Layout-Pytorch}}
    \item[*] RoPE \cite{su2023roformerenhancedtransformerrotary} implementation: \url{https://github.com/NormXU/ERNIE-Layout-Pytorch}
    \item[*] Batch Size: 48
    \item[*] Epochs: 10
    \item[*] Weight Decay: $1\times10^{-5}$
    \item[*] Dropout rate: 0.1
    \item[*] Optimizer: AdamW \cite{loshchilov2018decoupled}
    \begin{itemize}
        \item Learning Rate: $2\times10^{-5}$
        \item $\epsilon$: $1\times10^{-6}$
    \end{itemize}
    \item[*] LoRA:
    \begin{itemize}
        \item Rank: 256
        \item $\alpha$: 256
    \end{itemize}
    \item[*] All Parameters: 291,131,460
    \item[*] Trainable Parameters: 9,457,154 (3.25\%)
    \item[*] The ERNIE-Layout model was fine-tuned on an 1$\times$A100 machine for 12 hours.
\end{itemize}

\section{Implementation Details}
\label{sec:hyperparam}

The number of lookup tokens for Prompt Lookup Decoding is set to 3.
We used the implementation of the Transformers library for our experiments because it has the simplest implementation. We plan to move to vLLM \cite{kwon2023efficient} and SGLang \cite{zheng2024sglang} for faster speeds, but this will require more compatibility changes.
\section{Limitations}
Due to the limitations of the ERNIE layout model, our method chunks the edit sequence input to a maximum of 1024 tokens, but we have observed many pages exceeding this token count. 

Full-page formulas and tables cannot benefit from our method.
EditTrans may be less efficient in batch generation due to synchronization, and we are developing a more flexible approach.
\end{document}